\documentclass[pre,floatfix,twocolumn,showpacs]{revtex4}
\usepackage{amsmath}
\usepackage{amsfonts}
\usepackage{mathrsfs}
\usepackage{epsfig}
\usepackage{graphicx}
\usepackage{dcolumn}
\usepackage{expl3}
\usepackage{color}
\begin{document}
\title{
Using skewness and the
first-digit phenomenon to identify dynamical transitions in cardiac
models
}
\author{Pavithraa Seenivasan\,$^{1}$, Soumya Easwaran\,$^{1,*}$, 
S. Sridhar\,$^{1,2,*}$ and Sitabhra Sinha\,$^1$}
\affiliation{
${}^{1}$\mbox{The Institute of Mathematical Sciences, CIT Campus,
Taramani, Chennai 600113, India.}\\
${}^{2}$\mbox{Scimergent Analytics and Education Pvt Ltd, Adyar, Chennai,
India.}\\
${}^{*}$\mbox{These authors contributed equally to this work.}\\
}
\date{\today}
\begin{abstract}
Disruptions in the normal rhythmic functioning of the heart, termed as
arrhythmia, often result from qualitative changes in the excitation
dynamics of the organ. The transitions between different types of
arrhythmia are accompanied by alterations in the spatiotemporal
pattern of electrical activity
that can be measured by observing the time-intervals between
successive excitations of
different regions of the cardiac tissue. Using
biophysically detailed models of cardiac activity we show that
the distribution of these time-intervals exhibit a systematic change
in their skewness during such dynamical transitions. Further, the
leading digits of the normalized intervals appear to fit Benford's law
better at these transition points. This raises the possibility of
using these observations to design a clinical indicator for
identifying changes in the nature of arrhythmia.
More importantly, our results reveal an intriguing relation between
the changing skewness of a distribution and its agreement with
Benford's law, both of which have been independently proposed earlier
as indicators of regime shift in dynamical systems. 
\end{abstract}


\maketitle

\section{Introduction}
Many vital physiological processes are characterized by rhythmic activity,
ranging from the circadian clock regulating the daily sleep-wake
cycle to temporal patterns of respiration that occur over a
scale of seconds~\cite{Glass2001}. The periodic beating of the heart,
that results in constant circulation of oxygenated blood throughout the body,
is one of the most important of such naturally occurring oscillatory
phenomena in the body~\cite{Zipes2013}.
Certain types of disturbances in the cardiac rhythmicity, referred
to as arrhythmia, can severely impair the normal functioning of the
heart and in the most critical instances, can result in sudden cardiac
death~\cite{Winfree80}. Such ``dynamical diseases''~\cite{Glass77,Glass95},
i.e.,  diseases resulting from abnormal activity in an otherwise
intact physiological system, are a significant public health burden in
developed countries. For example, in the United States, diseases
of the heart constitute the leading cause of death (responsible for
about 25\% of all deaths), of which more than half can be classified as
sudden cardiac deaths~\cite{Heron2013,Hennekens1998,Zheng2001}. Even
in developing countries, in recent times heart disease
has overtaken other causes of death, e.g., sudden cardiac deaths
contributed to about 10\% of overall mortality in certain
regions in India, 
accounting for upto half of all cardiovascular-related 
deaths~\cite{Madhavan2011,Rao2012}.

Several studies have shown that early detection of onset of arrhythmia
resulting in prompt therapeutic intervention significantly improves the
chances of surviving such episodes~\cite{Gold2010,Travers2010}. 
Thus, developing methods
for identifying signs of impending arrhythmic events with potentially
serious consequences can significantly contribute towards reducing the
mortality rate due to sudden cardiac death. With this aim in view
there have been a number of attempts at applying time-series analysis
methods on cardiac activity data in order to extract robust indicators
of imminent instances of temporal irregularities in the heart.
However, the complexity of heart rate dynamics makes it difficult to
characterize and distinguish the temporal signatures of a healthy heart 
from a diseased one~\cite{Christini1995,Kurths1995,Goldberger2002,Cohen2002,Voss2009,Shiogai2010,Iatsenko2013}.

Most studies of cardiac time-series have focused on heart rate
variability 
as measured by temporal changes in the {\em R-R interval}, the duration between 
successive episodes of ventricular depolarization 
which triggers contraction of the lower chambers of
the heart.
Following the pioneering observations connecting decreased variance in R-R
intervals with higher mortality risk in patients suffering myocardial
infarction~\cite{Wolf1978,Kleiger1987}, it is now generally accepted
that healthy individuals have higher heart-rate variability compared
to those with diseased hearts~\cite{Lombardi2000}.
However, certain pathological conditions including cardiac arrhythmia
are seen to be extremely irregular~\cite{Costa2002}. In fact, it has been
observed that a transition
from tachycardia, i.e., abnormally rapid heart-rate, to fibrillation,
characterized by erratic muscle activity that prevents the heart from
pumping blood, is marked by a switch from relatively more periodic
activity to a highly irregular dynamical state~\cite{Garfinkel1997}. 
While the R-R
intervals in normal sinus rhythm appear to have almost as
unpredictable a nature as that seen during
fibrillation~\cite{Small2002}, it
has been suggested that the ``chaoticity'' during normal cardiac
activity arises through interaction of the heart
with the nervous system~\cite{Nakai2010}. In contrast, the
spatiotemporal chaos associated with fibrillation arises from
intrinsic instabilities in cardiac excitation 
dynamics~\cite{Weiss1999}. 

One possible route from tachycardia to fibrillation that has been
established through
extensive simulation studies of models of cardiac electrical activity
is the degeneration of reentrant spiral wave (corresponding to
tachycardia) to disordered, turbulent activity (characterizing
fibrillation) through spiral breakup~\cite{Fenton2002,Sridhar2013}. 
This dynamical
transition has been reproduced in a wide range of systems, from simple,
excitable media to biologically detailed models, underlining the
robustness of the scenario~\cite{Sinha2015}. 
Thus, the study of spatiotemporal dynamics in models of electrical activity
in cardiac tissue
provides another perspective to identify indicators for
an impending onset of possibly life-threatening arrhythmia.

In this paper, we focus on analyzing time-series data obtained from
spatially extended models of cardiac ventricular activity in which, 
by tuning specific physiological parameters, 
one can observe transitions to different dynamical
regimes representing various classes of arrhythmia. This allows us to
look for statistical signatures that can help in early detection of
arrhythmic episodes, where the observed patterns are
exclusively due to abnormal excitation activity that characterizes
such arrhythmia and unrelated to heart rate variability that arises
from the influence of the nervous system on the sinus node, 
the natural pacemaker of
the heart~\cite{Lombardi2011}. This study, therefore, provides a
benchmark against which analysis of ECG data obtained from clinical
studies can be compared, enabling distinction of statistical features
of arrhythmic time-series that are intrinsic to the dynamics of
heart muscle from those that are a result of
changes in the autonomic modulation of
cardiovascular function (achieved through dynamical balance between
sympathetic and parasympathetic effects~\cite{Purves2001}). 
As signature patterns (if they
exist) that indicate transitions from one dynamical regime of
cardiac activity to another may be masked by other effects in reality, 
establishing them through analysis
of the model output will allow us to look for them in data obtained
from experimental or clinical studies.

\begin{figure}
\begin{center}
\includegraphics[width=0.99\linewidth]{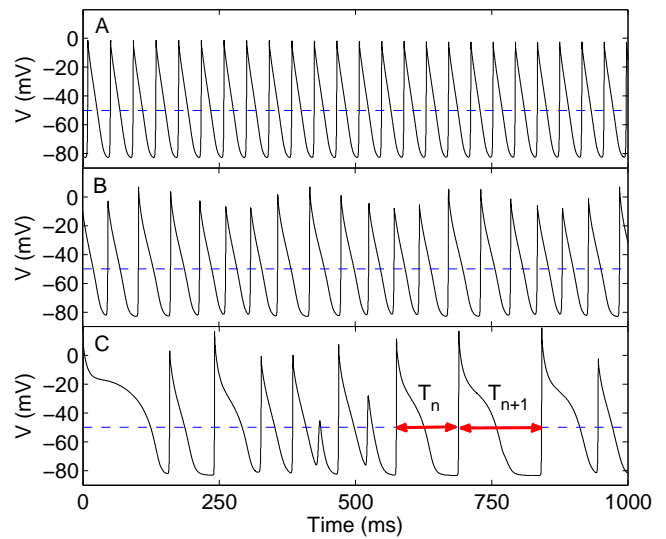}
\end{center}
\caption{Time-series of the 
transmembrane potential $V$ representing local excitation activity in a
two-dimensional LR1 model ($L = 400$) for different values of the
maximum Ca$^{2+}$ channel conductance $G_{si}$, viz., (A) 0.005, (B)
0.04 and (C) 0.065 mS
cm$^{-2}$. The distinct nature of the corresponding spatiotemporal 
dynamics, viz., rigid rotation of a spiral similar to monomorphic tachycardia (A), chaotic meandering of
spiral core representing polymorphic tachycardia (B), and 
spatiotemporal chaos indicating fibrillation (C), is visually apparent
in the varying pattern of intervals between successive
excitations. It is quantified in terms of the sequence of
time-intervals
$\{T_n\}$ ($n = 1, 2, \ldots$) between each pair of
consecutive local supra-threshold depolarizations (two such intervals
are shown in panel C using double-headed arrows). We consider a local
supra-threshold excitation event to have occurred when
$V$ exceeds $-50$ mV (broken line). 
}
\label{fig01}
\end{figure}

Here, for our statistical analysis, we have focused on the sequence of
time-intervals ${T}$ between successive excitations of ventricular muscle
cells (analogous to the R-R interval for ECGs) as a representative feature of
heart rate dynamics (Fig.~\ref{fig01}). 
An important result of our study is that the
distribution of these intervals exhibit clearly observable changes in
their moments - in particular, the skewness - around the onset 
of qualitatively distinct dynamical
behavior characterizing various arrhythmic episodes (represented by
the different panels in Fig.~\ref{fig01}).
Intriguingly, we also observe that at these transitions, the
distribution of the time-intervals appears to agree better with
Benford's law (BL), an empirically established feature of the
frequency distribution of leading digits of numbers occurring in 
many phenomena in various physical, biological and social
contexts~\cite{Hill1998,Sambridge2010,Friar2012,Nigrini1996}. 
Both variation in skewness~\cite{Guttal2008,Scheffer2009} and closer
agreement with BL~\cite{De2011} have
independently been suggested as indicators of regime shifts or phase
transitions in different systems. Our work not only finds both of
these signatures to be indicative of the onset of arrhythmic
behavior, but additionally suggests that these two phenomena (viz.,
increasing skewness and agreement with BL) may be related.

\section{Materials and Methods}
{\bf Model.} To simulate spatiotemporal excitation activity in cardiac
muscle under different physiological conditions, we have used a 
two-dimensional model of ventricular tissue having the generic form
\begin{equation} 
\frac{\partial V}{\partial t} = \frac{- I_{ion}(V,g_i)}{C_m}
+ D{\nabla}^2 V,
\label{eq1}
\end{equation}
where $V$ (mV) is the potential difference across a cellular membrane,
$C_m (= 1 \mu {\rm F} {\rm cm}^{-2})$ is the transmembrane
capacitance, $D$ is the diffusion constant
(~$=0.001 {\rm cm}^2 {s}^{-1}$ for the results reported in the paper), 
$I_{ion} (\mu{\rm A cm}^{-2})$ is the total current density 
through ion channels on the cellular membrane,
and $g_i$ describes the dynamics
of gating variables of different ion channels. The specific functional
form for $I_{ion}$ that we have focused on here is that of 
the Luo-Rudy I (LR1) model
which describes the ionic currents in a guinea pig ventricular
cell~\cite{Luo1991}:
\begin{equation*}
I_{ion} = I_{Na} + I_{K} + I_{K1} + I_{Kp} + I_{si} + I_{b},
\end{equation*}
where $I_{Na} = G_{Na} m^3 h j (V - 54.4)$ is the fast inward Na$^+$ 
current, $I_{si} = G_{si} d f (V-E_{si})$ is the slow inward Ca$^{2+}$
current where $E_{si} = 7.7 - 13.0287$ ln~([Ca$^{2+}$]$_i$) is the
reversal potential, dependent on the intracellular ion concentration
[Ca$^{2+}$], $I_{K} = G_K x x_1 (V+77.62)$, $I_{K1} = G_{K1}
K1_{\infty} (V+87.95)$ and $I_{Kp} = 0.0183 Kp (V+87.95)$
are three different types of K$^+$ current, 
and $I_{b} = 0.03921 (V+59.87)$ is a background current. The
currents are determined by ion channel gating variables
$m$, $h$, $j$, $d$, $f$ and $x$,
whose time evolution is described by 
ordinary differential equations of the form,
$d \epsilon /dt  = (\epsilon_{\infty} -
\epsilon)/\tau_{\epsilon}$,
where $\epsilon_{\infty}$ is the steady state value of $\epsilon$
(=$m$, $h$, $d$, $f$ and $x$) and
$\tau_{\epsilon}$ is
the corresponding time constant
obtained by fitting experimental data.
Parameter values used are as in Ref.~\cite{Luo1991}, except for $G_K$
which is set to $0.705$ mS/$\mu$F and $G_{si}$ that is varied to alter the
stability of spiral wave dynamics~\cite{Xie2001}.
We have explicitly verified that our results are not
sensitively dependent on model-specific details, viz., the description
of ion-channel dynamics, by observing qualitatively similar behavior
with another functional form for $I_{ion}$ described in the ten
Tusscher-Panfilov (TP06) model of a human ventricular cell~\cite{TP06}.

For numerical simulations, the two-dimensional spatially extended system 
is discretized on a 
grid of size $L \times L$ ($= 400$ for LR1 model and $1024$ for TP06 model) with a space step of $\delta x$ ($=
0.0225$ cm for LR1 model and $0.025$ cm for TP06 model). 
The equations are solved using a forward Euler method with time step 
$\delta t$ ($= 0.05$ ms for LR1 model and $0.02$ ms for TP06 model)
and a standard 5-point stencil for the
Laplacian describing the spatial coupling between the lattice elements.
No-flux boundary conditions are implemented at the edges.
%
The initial spiral wave state is
obtained by generating a broken wave front which then
dynamically evolves into a curved rotating wave.
The movement of the spiral wave core is obtained by tracing the
trajectory of intersection points of iso-contour lines
for a pair of dynamical variables of the model, viz., $V$ and
$h$ in the LR1 model~\cite{Barkley1990}. 

{\bf Inter-spike interval time-series.}
We analyze the statistical properties of the time-intervals between
successive excitations (i.e., depolarization) at specific locations
in the simulated cardiac tissue. 
Each point in the simulation 
domain is considered to be excited if the
corresponding transmembrane potential $V$ crosses a threshold value
(set equal to $-50$ mV here, although our results are robust
with respect to the choice of threshold) from below, i.e., from a
hyperpolarized state.
The time-interval $T$ between two such successive crossings of the
threshold is recorded for constructing the data-set, 
values being sampled from 400 equally spaced points arranged in a regular
grid   
on the simulation domain (for the LR1 model). 
Typically time series of $5$ s total duration were used for our analysis.
From the data-sets obtained at different values of $G_{si}$, the corresponding 
distributions for $T$ are obtained and the moments calculated,
including mean $\mu$, standard deviation $\sigma$, and skewness, the
latter being measured by
the Pearson's moment coefficient of
skewness defined as
$\gamma = E[(X-\mu)/\sigma)^3]$.
Using other measures for 
the skewness did not qualitatively alter our results.
The time-interval distributions obtained for different parameters
are also tested for the degree of agreement with Benford's Law.

{\bf BL and Benford distribution.}
Named after the American physicist F. Benford who made the first-digit
phenomenon widely known,
BL had been noticed in numbers associated with a variety of
natural (as well as social) phenomena as far back as in 1881 
by the astronomer S. Newcomb.
According to this empirical law, numbers beginning with 1 or 2 occur
more often than those beginning with 8 or 9~\cite{Fewster2009}. 
Specifically, the probability 
of the first or
leading digit of such numbers being $i$ ($i = 1, 2, \ldots 9$) is
given by the {\em Benford distribution}:
\begin{equation*}
P(i) = \log_{10}\left(1+\frac{1}{i}\right).
\end{equation*}
The reason for the ubiquity of this distribution has been connected to
its scale-invariance and base-invariance~\cite{Hill1995}. 
Thus, if indeed there is a
universal principle underlying the distribution of the leading digits of
numbers which is independent of the units in which the numbers are
measured or the number base used, then the BL follows.
Simple mathematical arguments have been used to show that any
distribution of numbers arising from natural processes that spans
several orders of magnitude and is reasonably smooth will obey
BL~\cite{Fewster2009}. The Benford distribution, as mentioned earlier,
is seen in many empirical data-sets, including those arising in a
biological context, such as, the distribution of open reading frames
in prokaryotic and eukaryotic genomes~\cite{Friar2012}.
Dynamical systems, such as those describing the molecular
dynamics of fluids or certain chaotic systems, also exhibit BL 
in the numbers expressing coordinates of the generated 
trajectories~\cite{Tolle2000,Snyder2001}.
More recently, BL has been used as a signature for detecting phase 
transitions in a quantum system~\cite{De2011}.

In order to compare the distribution of the intervals $T$ between
successive
excitations with that expected from BL, we first obtain a set of
normalized time-intervals $t$
by subtracting the minimum value of the series from all intervals $T$
and dividing by the range, i.e., $t =$
[$T-$min($T$)]/[max($T$)$-$min($T$)].
The leading digits $i$ of the normalized intervals $t$ are then
extracted
as $i = \lfloor |t|/10^{\lfloor \log_{10}(|t|)\rfloor}\rfloor$, where
$| z |$ and
$\lfloor z \rfloor$ indicate the absolute value of $z$
and the largest integer not greater than $z$ respectively.
The distribution of $i$ is then tested for agreement with 
BL using statistical tests for goodness of fit.

{\bf Statistical tests for goodness of fit with BL.}
The goodness of fit between the two distributions (the empirical and
that predicted by BL) is measured by a two-sample
Kolmogorov-Smirnov test.
We have used the function $kstest2$ implemented in {\em MATLAB} which
returns a test decision for the null hypothesis
that both the sets are from the same distribution, along with a
$p$-value and the KS test statistic $k$ describing the
degree of deviation from BL.
In our study the test statistic is the comparison parameter
$$k = \smash{\displaystyle\max_{i}} (F_c (i) - B_c (i)),$$
which measures the maximum distance
between
the two cumulative distributions, $F_c (i)$ and $B_c (i)$, of the
leading digits $i$ of normalized time intervals and that expected from
BL, respectively.
A lower value of $k$ implies closer agreement with
the Benford distribution.

Apart from the KS test, we have also used the Pearson's
chi-squared test to confirm the compliance of the empirical
distributions with BL.
The test statistic
$$\chi^2 = \sum\limits_{j=1}^{n} \frac{(F_{j} - B_{j})^2}{B_j},$$
quantifies the total magnitude (over all $n$ entries of the
empirical time-series) of the difference between the two
probability distributions, $F$ and $B$, for the
leading digits of the normalized time intervals and that expected from
BL, respectively. As for the KS test,
a lower value of $\chi^2$ implies closer agreement
with the Benford distribution.

\section{Results}
To identify the statistical signatures characterizing dynamical
transitions to different types of arrhythmia, we systematically
explore the spatiotemporal dynamics of the model systems in different
parameter regimes.
The nature of the excitation activity is varied in
a controlled manner by changing the kinetic properties of an ion
channel, viz., increasing the maximum Ca$^{2+}$ channel
conductance $G_{si}$ for the LR1 model 
(keeping all other model parameters unchanged)
which is known to result in a succession of dynamical
transitions~\cite{Qu2000} as seen in Fig.~\ref{fig01}.
For the TP06 model, the maximum conductance
$G_{pCa}$ of the sarcolemmal pump current $I_{pCa}$ is increased
that eventually results in spiral breakup leading to spatiotemporal
chaos~\cite{TP06}.

\begin{figure*}
\begin{center}
\includegraphics[width=0.99\linewidth]{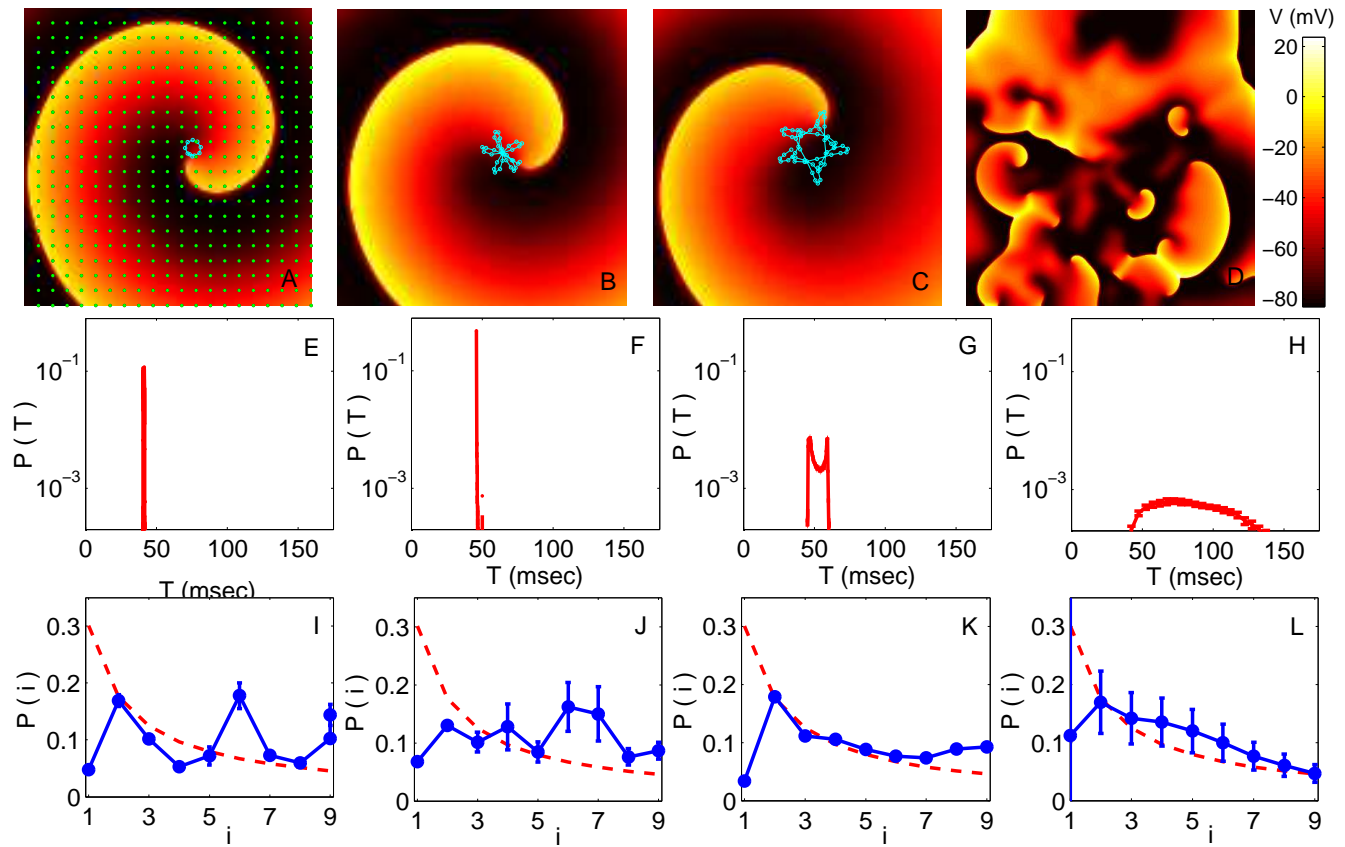}
\end{center}
\caption{
(A-D) Pseudocolor images of spatiotemporal activity
(measured in terms of transmembrane
potential $V$) for the two-dimensional LR1 model ($L = 400$) showing
the different dynamical regimes obtained by increasing the maximum
Ca$^{2+}$ channel conductance $G_{si}$ (expressed in units of mS
cm$^{-2}$).
The successive panels represent a spiral wave undergoing (A) stable
rotation ($G_{si} = 0.005$), (B) quasiperiodic meandering ($G_{si} =
0.025$) and (C) chaotic meandering ($G_{si} = 0.04$). Further increase
of $G_{si}$ results in breakup of the spiral wave leading to
(D) spatiotemporal chaos ($G_{si} = 0.065$).
The trajectory of the spiral
core (the tip of the spiral wave, defined to be a phase singularity)
for a duration of 500 ms is indicated in all panels except for the one corresponding to chaotic
activity where there is a large multiplicity of singularities.
(E-H) The probability distribution of time intervals $T$ between
successive supra-threshold activations of a local region corresponding
to the
dynamical regimes indicated in panels (A-D) respectively. 
(I-L) Probability distribution of the leading digits $i$ of the
normalized
time intervals between successive supra-threshold activations of a
local region corresponding to the
dynamical regimes indicated in panel (A-D) respectively.
The broken curve indicates the distribution predicted by Benford's
law.
Each distribution in panels E-L is obtained by
averaging over data collected from many spatial positions in the
simulation domain (indicated by points in
panel A) and also over
several realizations, with error bars indicating the 
standard deviation.
}
\label{fig02}
\end{figure*}
Representative images of the spatiotemporal dynamics (with 
LR1 model ion-channel kinetics) in the different
regimes
are shown in the top row of Fig.~\ref{fig02}, with the trajectory of
the spiral core (traced in the first
three panels using a light color) exhibiting characteristic changes in
its qualitative nature.
Starting from an initial state
characterized by a rotating spiral wave, for small values of
the conductance (e.g. $G_{si} = 0.005$) we observe rigid rotation
with the core moving around an approximately
circular trajectory (Fig.~\ref{fig02}~A), which corresponds to the clinical
phenomenon of monomorphic tachycardia.
This gives way to meandering at higher values of
$G_{si}$ ($\simeq 0.025$, see Fig.~\ref{fig02}~B), followed by the
appearance of chaotic
meandering around $G_{si} = 0.04$ (Fig.~\ref{fig02}~C) and finally the
breakup of spiral
waves leading to spatiotemporal chaos, representative of fibrillation,
for values of $G_{si} >  0.055$
(Fig.~\ref{fig02}~D).

The panels in the middle row of Fig.~\ref{fig02} show
the probability distribution of
time intervals between successive excitations, $T$, for the
$G_{si}$ values corresponding to the panels in the top row. We observe
that the range of these intervals become broader at larger $G_{si}$
values as the dynamics of the spiral wave becomes more complex.
A very narrow range of intervals is dominant at low $G_{si}$, as
expected for a rigidly rotating spiral wave having a characteristic
period of rotation (Fig.~\ref{fig02}~E).
With increased meandering of the core, however,
the time interval between successive excitations of a local region
becomes more irregular, which is manifested as a broader distribution
of $T$ (Fig.~\ref{fig02}~F).
As the spiral core trajectory becomes even more complex,
covering a larger portion of the simulation domain, we see that the
distribution not only widens further but also develops multiple peaks
at the extremities (Fig.~\ref{fig02}~G). Finally, following breakup
and spatiotemporal
chaos, the time between successive excitations become essentially
random in character with a distribution that spans a relatively large
range of $T$ (Fig.~\ref{fig02}~H).

To see how closely the dynamical process follows the Benford
distribution in the different regimes, in the
bottom row of Fig.~\ref{fig02} we show the probability distributions
of the leading digits $i$ of the normalized intervals $t$.
It is evident that the distribution of $i$ moves closer to the form
expected for the Benford distribution
(indicated by a broken curve) at larger values of $G_{si}$.
In fact, the empirical distribution shows the best agreement with BL
in the spatiotemporally chaotic state corresponding to $G_{si} =
0.065$ (Fig.~\ref{fig02}~L), which is consistent with the
corresponding time interval
distribution being exponential in nature - as it is known that values
distributed exponentially follow BL.
We see that that this
distribution of leading digits $i$ is closest to BL at the
transition points corresponding to chaotic meandering ($G_{si} =
0.040$) and spiral breakup ($G_{si} = 0.065$).

\begin{figure*}
\begin{center}
\includegraphics[width=0.99\linewidth]{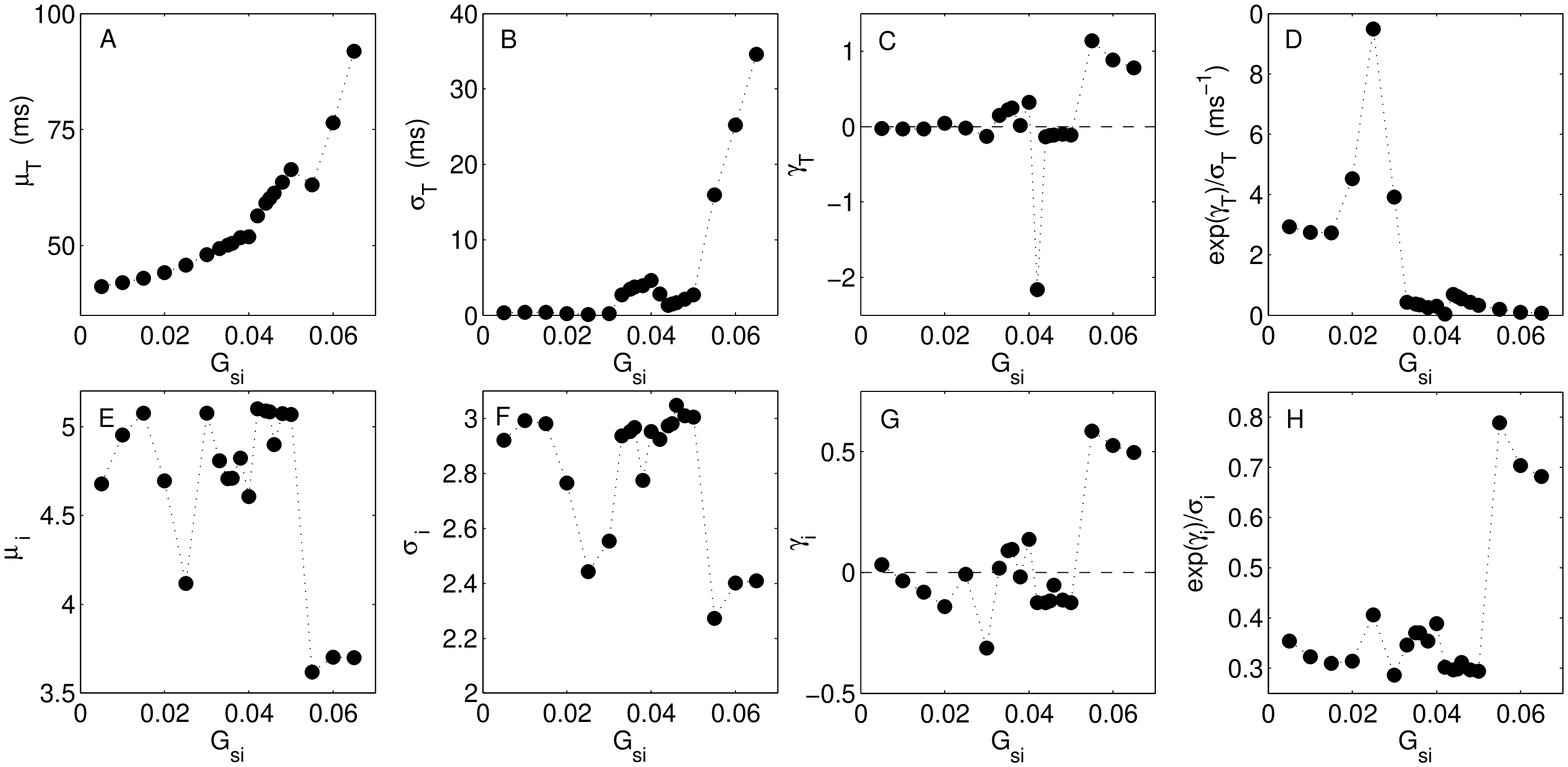}
\end{center}
\caption{
Analysis of various moments for the distributions of
the time intervals $T$ between successive supra-threshold activations 
of a local region (A-D) and that of the leading
digits $i$ of the normalized time intervals (E-H)
as a
function of the maximum Ca$^{2+}$ channel conductance $G_{si}$
in the two-dimensional LR1 model.
Variation in (A) the mean $\mu_T$, (B) standard deviation $\sigma_T$,
(C) skewness $\gamma_T$ measured in terms of the Pearson's moment
coefficient and (D) the derived quantity $exp(\gamma_{T})/\sigma_{T}$,
correspond to the distribution of the time intervals $T$, while the
variation shown for (E) the mean $\mu_{i}$, (F) standard deviation
$\sigma_i$, (G) skewness $\gamma_i$ (again measured in terms of the
Pearson's moment coefficient), and (H) the quantity 
$exp(\gamma_{i})/\sigma_{i}$, correspond to the distribution of the
leading digits $i$.
Large changes in both the skewness measures
($\gamma_T$ and $\gamma_i$), and to an extent, the dispersion measures
($\sigma_T$ and $\sigma_i$) correspond
to successive dynamical transitions between rigid rotation,
quasiperiodic meander and chaotic meander of the spiral core, finally
giving rise to spatiotemporal chaos resulting from spiral breakup.
The measure $exp(\gamma_{i})/\sigma_{i}$ combines the
information obtained from the variation of standard deviation and skewness, 
enabling it to indicate some of the
dynamical transitions more clearly.
The linear correlation coefficient between the
skewness of $T$ and that of $i$ is $r_{\gamma} = 0.67$
($p$-value$=0.001$). 
}
\label{fig03}
\end{figure*}
To understand the nature of the distributions in the different
dynamical regimes better, we show how the moments of the distribution
for the time intervals $T$ and that for the corresponding leading
digits $i$ of the normalized intervals vary with increasing $G_{si}$
(Fig.~\ref{fig03}). We observe that the mean value of the interval
between successive excitations steadily rises with $G_{si}$ as the
complexity of the spiral core trajectory increases excepting for a
small dip around $G_{si} = 0.055$ which is the point of transition to
spiral breakup (Fig.~\ref{fig03}~A). The dispersion of the $T$
distribution, measured by its standard deviation $\sigma_T$
(Fig.~\ref{fig03}~B) shows a
similar increasing nature with $G_{si}$ although, around $G_{si} =
0.04$, where a transition occurs from quasiperiodic to chaotic
meandering of the spiral core, there is a small decrease.
The skewness $\gamma$ of the distribution is the
most
informative of all the moments considered here, as it shows large
deviations from zero only around critical values of $G_{si}$
associated with transitions between different dynamical regimes. In
particular, we notice peaks in the skewness at $G_{si} = 0.025$,
$0.04$ and $0.055$ which correspond to transition to quasiperiodic
meandering, chaotic meandering and spiral breakup, respectively
(Fig.~\ref{fig03}~C).
In order to make the relation between the different moments and
the dynamical transitions even more clear, we have also shown the
nature of variation of a derived quantity, exp($\gamma_T$)/$\sigma_T$,
as
a function of $G_{si}$. It can potentially be used as a statistical
indicator for the onset of certain types of arrhythmia that may be
hard to detect by observing the skewness alone. We see from
Fig.~\ref{fig03}~(D) that the measure amplifies the signal indicating a
transition close to $G_{si} = 0.025$ where the spiral begins to
noticeably meander.

When we observe the corresponding moments for the distribution of
leading digits $i$ as a function of $G_{si}$, we note that both
the moments $\mu_i$ and $\sigma_i$
(Fig.~\ref{fig03}~E-F) have very similar
nature of variation, viz., both exhibit dips around the values
of $G_{si}$ at which the different dynamical transitions occur.
In contrast, the skewness exhibits an almost opposite nature,
with peaks occurring at the transition points (consistent with
increasing skewness of the $T$ distribution at these values).
This indicates that at these points the distribution comes close to
the form expected from
BL, as the latter is associated with positively skewed
distributions.
The derived quantity exp($\gamma_i$)/$\sigma_i$ conserves this
pattern, showing increased values at these points.
We find that the skewness of $T$ and that of $i$ are correlated, the
linear correlation coefficient between $\gamma_{T}$ and $\gamma_{i}$
being $r_{\gamma} = 0.67$ ($p$-value$=0.001$). This indicates an
inter-dependence between the variations in skewness of the time-interval
distribution and that of the
leading digit distribution, that occur at different dynamical
transitions.

\begin{figure*}
\begin{center}
\includegraphics[width=0.99\linewidth]{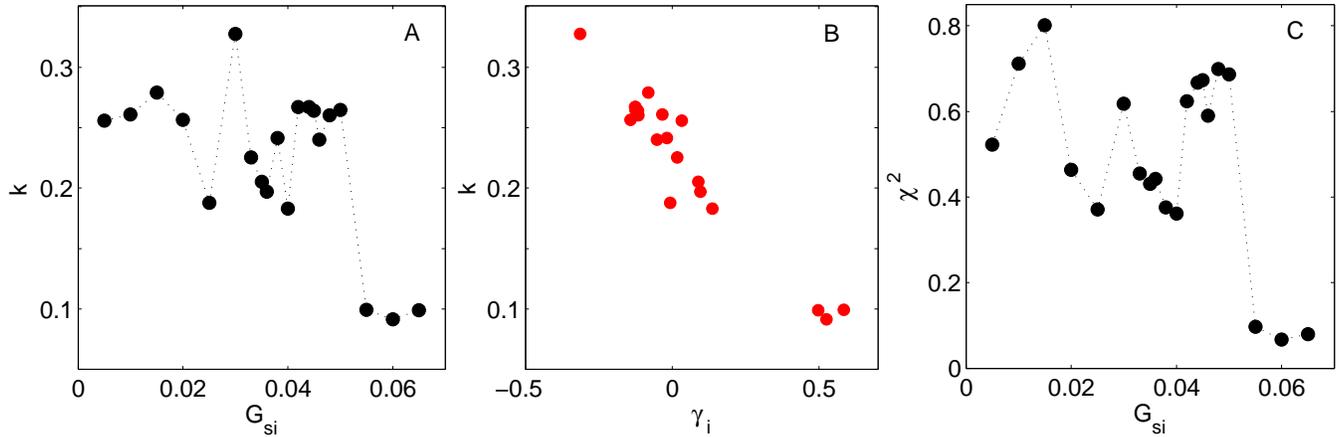}
\end{center}
\caption{
(A) Deviation of the distribution of leading digits $i$ of the
normalized time intervals $t$ from Benford's law in the
two-dimensional LR1 model measured in terms of
the Kolmogorov-Smirnov test statistic $k$ and shown as a function of
the maximum Ca$^{2+}$ channel conductance $G_{si}$.
(B) There is a strong negative correlation ($r =
-0.96$, $p$-value = $10^{-12}$) between $k$ and the skewness of
the leading digits $\gamma_i$.
(C) The error in using BL for describing the empirical
distribution of leading digits $i$ of the
normalized time intervals $t$ is measured using
Pearson's chi-squared test and shown as a function of $G_{si}$.
Agreement with BL improves whenever there is a dynamical
transition, as seen by dips in $k$ and the $\chi^2$
test statistic for values of $G_{si}$ where successive transitions
between rigid rotation, quasiperiodic meandering and chaotic meandering
of the spiral core and spatiotemporal chaos occur.
}
\label{fig04}
\end{figure*}
\begin{figure*}
\begin{center}
\includegraphics[width=0.99\linewidth]{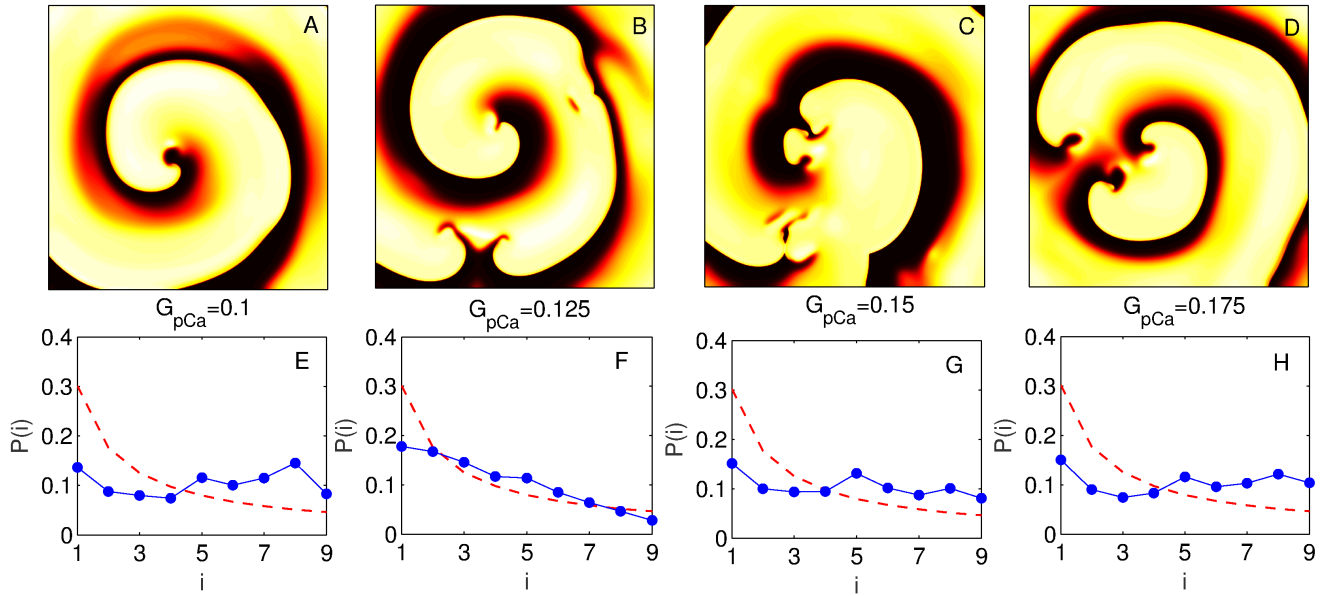}
\end{center}
\caption{(A-D) Pseudocolor images of spatiotemporal activity (measured in terms of transmembrane potential $V$) for the
two-dimensional TP06 model ($L = 1024$) showing the different dynamical regimes obtained by increasing the maximum conductance $G_{pCa}$
of the sarcolemmal pump current (expressed in units of nS pF$^{-1}$). The successive panels represent
a single spiral wave at $G_{pCa} = 0.1$ (A), the initiation of spiral breakup at $G_{pCa} = 0.125$ (B) and successive states leading to spatiotemporal chaos   
at $G_{pCa} = 0.15$ (C) and $G_{pCa} = 0.175$ (D). (E-H) Probability distribution of the leading digits $i$ of the normalized time intervals between successive supra-threshold activations of a local region corresponding to the dynamical regimes indicated in panel (A-D) respectively. The broken curve indicates the distribution according to Benford’s law. The analysis shown in panels E-H is performed for data obtained from a grid of $225$ equally spaced points in the two-dimensional medium.
}
\label{fig05}
\end{figure*}
\begin{figure}
\begin{center}
\includegraphics[width=0.99\linewidth]{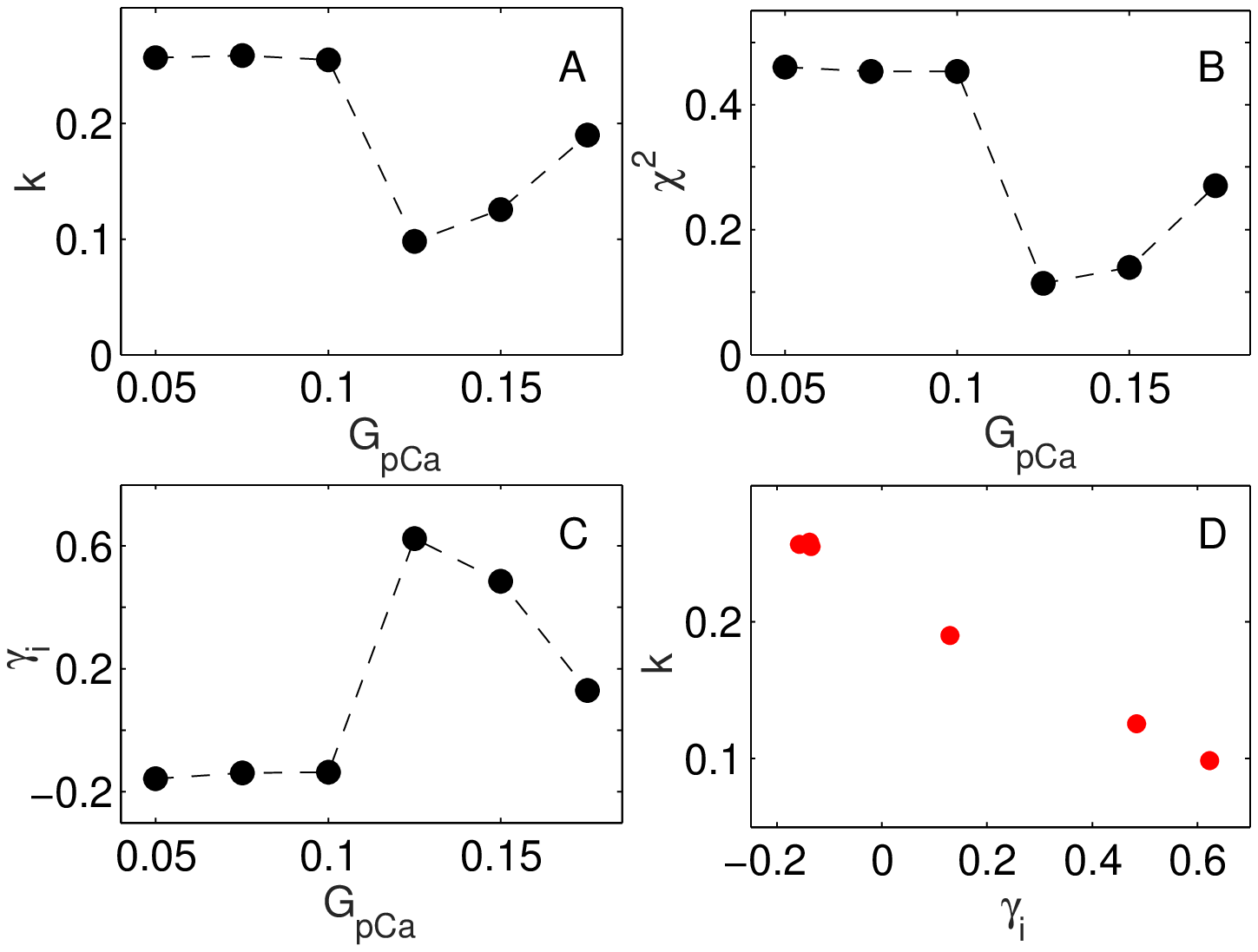}
\end{center}
\caption{(A-B) Deviation of the distribution of leading digits $i$ of the normalized time intervals $t$ from BL in the two-dimensional
TP06 model measured
in terms of (A) the Kolmogorov-Smirnov test statistic $k$ and 
(B) Pearson's Chi-squared test statistic $\chi^2$,
shown as a function of the maximum conductance $G_{pCa}$
of the sarcolemmal pump current (expressed in units of nS pF$^{-1}$).
Agreement with BL is highest around the value of $G_{pCa} = 0.125$
where breakup of the spiral wave is initiated,
as seen by dips in $k$ and the $\chi^{2}$ 
test statistic.
(C) Skewness $\gamma_i$ in the distribution of the leading digits of the normalized time interval, with the peak occurring 
at $G_{pCa} = 0.125$ corresponding to the spiral breakup transition point.
(D) There is a strong negative correlation ($r = -0.99$, $p$-value = $10^{-7}$) between $k$ and the skewness of the leading
digits $\gamma_i$.}
\label{fig06}
\end{figure}

To quantify how closely the system obeys BL in the different dynamical
regimes, we show the results of different statistical tests for
goodness of fit between the empirical and Benford distributions.
Fig~\ref{fig04}~(A) shows the Kolmogorov-Smirnov (KS) test statistic
as a function of the Ca$^{2+}$ channel conductance $G_{si}$ which
clearly indicates that the distribution of leading digits follow BL
most closely, indicated by dips in the test statistic,
at the values of $G_{si}$ characterizing the different dynamical
transitions, viz., $G_{si} = 0.025$, $0.04$ and $0.055$ ($p$-values
for the statistic are effectively zero for all $G_{si}$).
Note that low values of the KS test statistic (i.e., better agreement
with BL)
are associated with high positive skewness of the distribution of
leading digits of the normalized intervals.
This is underlined by the strong negative correlation between the test
statistic $k$ and the skewness $\gamma_i$ (see Fig.~\ref{fig04}~B),
having a linear correlation coefficient $r = -0.96$ ($p =
10^{-12}$).
As mentioned earlier,
this is consistent with the fact that BL is associated with
distributions having high positive skewness.
Fig.~\ref{fig04}~(C) shows the result of another statistical test,
viz., Pearson's Chi-squared test, with the lowest values of Pearson's
error
- corresponding to better agreement with BL - occurring at the
same values of $G_{si}$ where the dynamical
transitions occur.
The points at which the empirical distribution best matches BL are
seen to be consistent for the two tests (the dips of the two test
statistics occurring at the same values).

The change in skewness and the agreement with BL around the transition 
to  spiral breakup shown here do not appear to be model specific. We have explicitly verified that qualitatively similar results can be obtained by
using the TP06 model for a human ventricular cell. The top row of Fig.~\ref{fig05} shows images representative of the spatiotemporal dynamics
of a two-dimensional medium in different regimes as the maximum value of conductance $G_{pCa}$ of the sarcolemmal pump current $I_{pCa}$ is increased.
For small values of $G_{pCa}$ a single spiral wave is seen to rotate
stably in the medium, 
until around $G_{pCa} = 0.125$ nS pF$^{-1}$ it undergoes breakup and degenerates into spatiotemporal chaos for higher values of $G_{pCa}$.
The panels in the bottom row show the distributions of the leading
digits for the normalized intervals at the corresponding values of
$G_{pCa}$ indicating that the fit with Benford distribution is closest
during the dynamical transitions.
This is rigorously shown in Fig.~\ref{fig06} where the results of different statistical tests 
for goodness of fit are given. Panels (A) and (B) show that both the KS and Pearson's 
Chi-squared tests point to a better agreement with BL at the transition point
$G_{pCa} = 0.125$ nS pF$^{-1}$. As with the results for LR1 model reported earlier, the transition is also associated with a large deviation in the skewness $\gamma_i$ (Fig.~\ref{fig06},~C) and
a strong negative correlation between the test statistic $k$ and the skewness (Fig.~\ref{fig06},~D)
with a linear correlation coefficient $r = -0.99$ ($p= 10^{-6}$). 

To summarize the results, around the parameter values
where the transitions between dynamical regimes representative of
different types of cardiac arrhythmia occur, we observe both higher
positive skewness and closer agreement with BL (as indicated by
statistical tests). We note that both increased skewness and better
match with BL have independently been suggested earlier as possible
signatures for dynamical transitions, although not in the context of
physiology or clinical applications.
Apart from the potential utility of this
observation for devising robust indicators of the onset of
life-threatening disturbances in the cardiac rhythm, it suggests a
deep relation between the appearance of BL
in natural phenomena and the degree of skewness in the
distributions of the underlying variables.

\section{Discussion}
Statistical analysis of data (in particular, ECG) that is representative of
cardiac functionality can provide us with effective signatures for the
detection of arrhythmia at an early stage. 
Despite such analyses, certain kinds of arrhythmia fail to
get detected merely due to the complexity involved. 
Part of the difficulty lies in cardiovascular activity being a
joint outcome of intrinsic spatiotemporal excitation dynamics in heart
muscle and modulation of the sinus node by the sympathetic and
parasympathetic nervous system. Here we have studied
biophysically detailed models of ventricular activity
to infer
signatures of dynamical transitions characterizing the onset of different
kinds of arrhythmia. This makes it possible to disentangle the effects of
intrinsic excitation dynamics in the heart from the influence of the
nervous system. In principle, it
allows identification of patterns that may alert one to
impending harmful disruptions in the rhythmic activity of the heart,
but which could be masked in the ECG signal by autonomic modulation effects.
Our results show that as the spatiotemporal excitations in the
ventricles become more disordered, leading to phenomena identified
with sustained monomorphic and polymorphic tachycardia as well
as the onset of fibrillation, these transitions are marked by
characteristic changes in statistical moments associated with the
distribution of inter-activation intervals. In addition, the leading
digits of these intervals show a closer agreement with BL at the
transition points. Our result can potentially be 
applied in augmenting algorithms
used in implanted devices (ICDs) for detecting transitions to possible 
life-threatening arrhythmia so as to initiate a program of 
treatment~\cite{Schuckers1998}. Thus, when continual
monitoring of heartbeat time-series shows either increased skewness or
a closer agreement to BL, it may signal
a transition in the dynamical state of the
heart so that suitable pacing therapy can be started.
However, for such an application, our observations would need to
be validated in ECG recordings made during tachycardia and onset of fibrillation in live animals. Such validation is necessary in view
of 
the limitations of the present study, involving
as it does two-dimensional monodomain models of homogeneous 
cardiac tissue.  

Examining the passage from normal cardiac activity to different
arrhythmic regimes from the perspective of phase transitions can provide
novel insights, as has been pointed out by several earlier studies.
For instance,  
power-law behavior, which characterize critical phenomena in physical
systems, have been reported in
R-R interval fluctuations and are seen to be
remarkable predictors of arrhythmic death, with a steeper negative
slope of the power spectra (in log scale) 
clearly distinguishing a diseased heart
from a healthy one~\cite{Bigger1996}. 
More recently, it has been shown that phase 
transition-like dynamics is exhibited by healthy human heart rate,
indicated by long range correlations which is a hallmark of
criticality~\cite{Kiyono2005}. In contrast, the dynamics of an
abnormal heart
rate reveals significant digression from critical behavior.
In addition, scale invariance, which is seen in systems
close to critical point, has been shown
to be indicative of a healthy heart - with its absence being a 
statistical feature that can alert us about 
pathological conditions~\cite{Kiyono2004}. 
Our results provide yet another connection between onset of arrhythmia
and phase transitions by showing that sharp changes in the skewness of
the distribution of dynamical variables, that has previously been 
associated with dynamical transitions in other
systems~\cite{Guttal2008,Scheffer2009}, can potentially act as a 
robust indicator of transitions between monomorphic tachycardia,
polymorphic tachycardia and fibrillation in the heart.

As mentioned earlier, the appearance of BL has also been linked to phase
transitions in physical systems~\cite{De2011}. As the Benford
distribution follows from requirement of scale invariance of the
underlying numbers~\cite{Hill1995}, it is tempting to connect this
with the scale invariance of distribution of dynamical variables
associated with critical points at which transitions occur.
We observe from our results that parameter regimes that give rise to
relatively broad distributions of the time-intervals (as is the
situation for spatiotemporally chaotic states) show better
agreement with BL than those associated with highly confined
distributions (as in the case of a rigidly rotating spiral). However,
the occurrence of chaos in dynamical systems by itself does not
guarantee that BL will be obeyed~\cite{Tolle2000}. Thus, it appears that the
appearance of Benford distribution is more closely associated with the
onset of dynamical transitions rather than the specific nature of the
dynamical states on either side of the transition point.
We also note that distributions that follow BL are, in general,
associated with high skewness~\cite{Scott2001}.
This suggests that the highly skewed nature of distributions during dynamical
transitions and the
observation of better agreement with BL at those points may not be
independent of each other. 
Thus, 
our results imply that the increased skewness
associated
with regime shifts and the appearance of Benford distribution during phase
transitions - which have been reported earlier in
different contexts - are, in fact, related.

\begin{acknowledgments}
We are grateful to Ujjwal Sen who has been
instrumental in getting us interested in the possibility of observing
BL in cardiac phenomena.
We would also like to thank Indrani Bose, K. Chandrashekar and 
Shakti N. Menon for several helpful suggestions. 
This research was supported in part by the IMSc Complex Systems (XII
Plan) Project funded by the Department of Atomic Energy, Government of
India.
We thank the HPC facility at IMSc for
providing access to ``Satpura'' which is partly funded by the
Department of Science and Technology (DST), Government of India
(Grant No. SR/NM/NS-44/2009).
\end{acknowledgments}


\end{document}